\documentclass{article}
\usepackage{hiph-art}
\input epsf
\volnumber{0} \issuenumber{1} \edyear{2005}                              
\frompage{000} \topage{000}                                              
\recrevdate{16 November 2005}                                            
\def\alphas{\alpha_{\rm s}}
\def\Mrow{Mr\'owczy\'nski}
\def\v{{\bf v}}
\def\D{{\bf D}}
\def\k{{\bf k}}
\def\p{{\bf p}}
\def\E{{\bf E}}
\def\B{{\bf B}}
\def\x{{\bf x}}
\def\rightmark{QGP Instabilities and Implications}
\def\leftmark{Guy D.\ Moore}
 
\title{Numerical Studies of QGP Instabilities and Implications} 
\authors{ 
{Guy D.~Moore$^{1}$ %
\index{Moore, G.\ D.} 
}\\[2.812mm]
{\normalsize
\hspace*{-8pt}$^1$    Department of Physics,
    McGill University, 3600 rue University\\
    Montr\'eal QC H3A 2T8, Canada
}}
 
\abstract{Because of the flat initial shape of the QGP in a heavy ion
collision, the momentum distribution becomes anisotropic after a
short time.  This leads to plasma instabilities, which may help explain
how the plasma isotropizes.  We explain the physics of instabilities and
give the latest results of numerical simulations into their evolution.
Nonabelian interactions cut off the size to which the soft unstable
fields grow, and energy in the soft fields subsequently cascades towards
more ultraviolet scales.  We present first results for the power
spectrum of this cascade.
}
\keyword{Plasma, Weibel instability, Thermalization, Heavy ion collisions}

\PACS{}
 
\makeindex
\begin{document}
 
\maketitle

\section{Introduction}\label{intro}

\begin{figure}
\centerline{\hfill\epsfxsize=0.4\textwidth\epsfbox{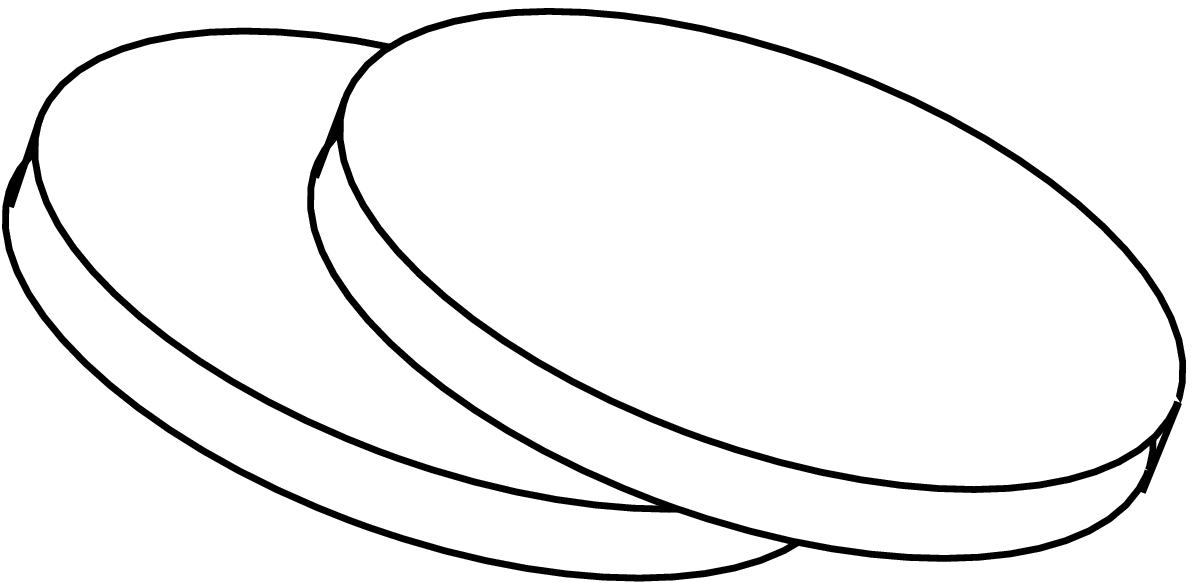}\hfill
\epsfxsize=0.2\textwidth\epsfbox{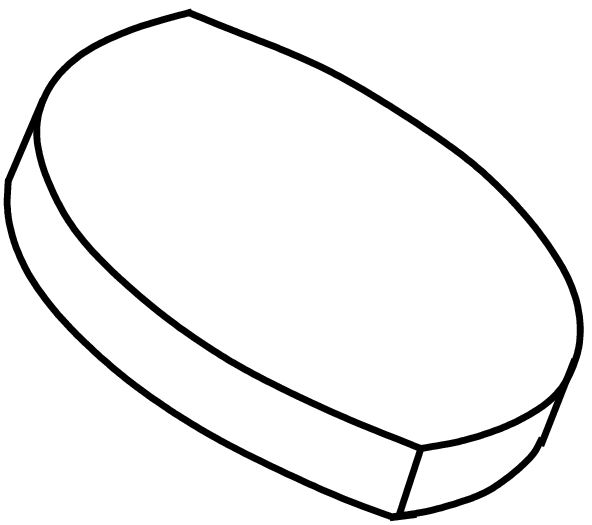}\hfill}
\caption{\label{fig1}Left:  pancake shaped nuclei just before
colliding.  Right:  resulting ``flat almond'' shaped region of
quark-gluon plasma.}
\end{figure}

When relativistic nuclei collide, they leave behind a plasma of quarks
and (mostly) gluons which starts out in a flat pancake shaped
region of space.  This simply reflects the fact that the nuclei are
approximately spherical in their rest frames, but the Lorentz boost of
their motion compresses them along the beam axis into nearly flat
sheets.  Further, if the collision is not perfectly head-on but occurs
at a finite impact parameter (as is usually the case), the initial
region is not circular in the transverse plane, either; it looks instead
like a ``flat almond,'' as depicted in Figure \ref{fig1}.

\begin{figure}
\centerline{\hfill \epsfxsize=0.42\textwidth\epsfbox{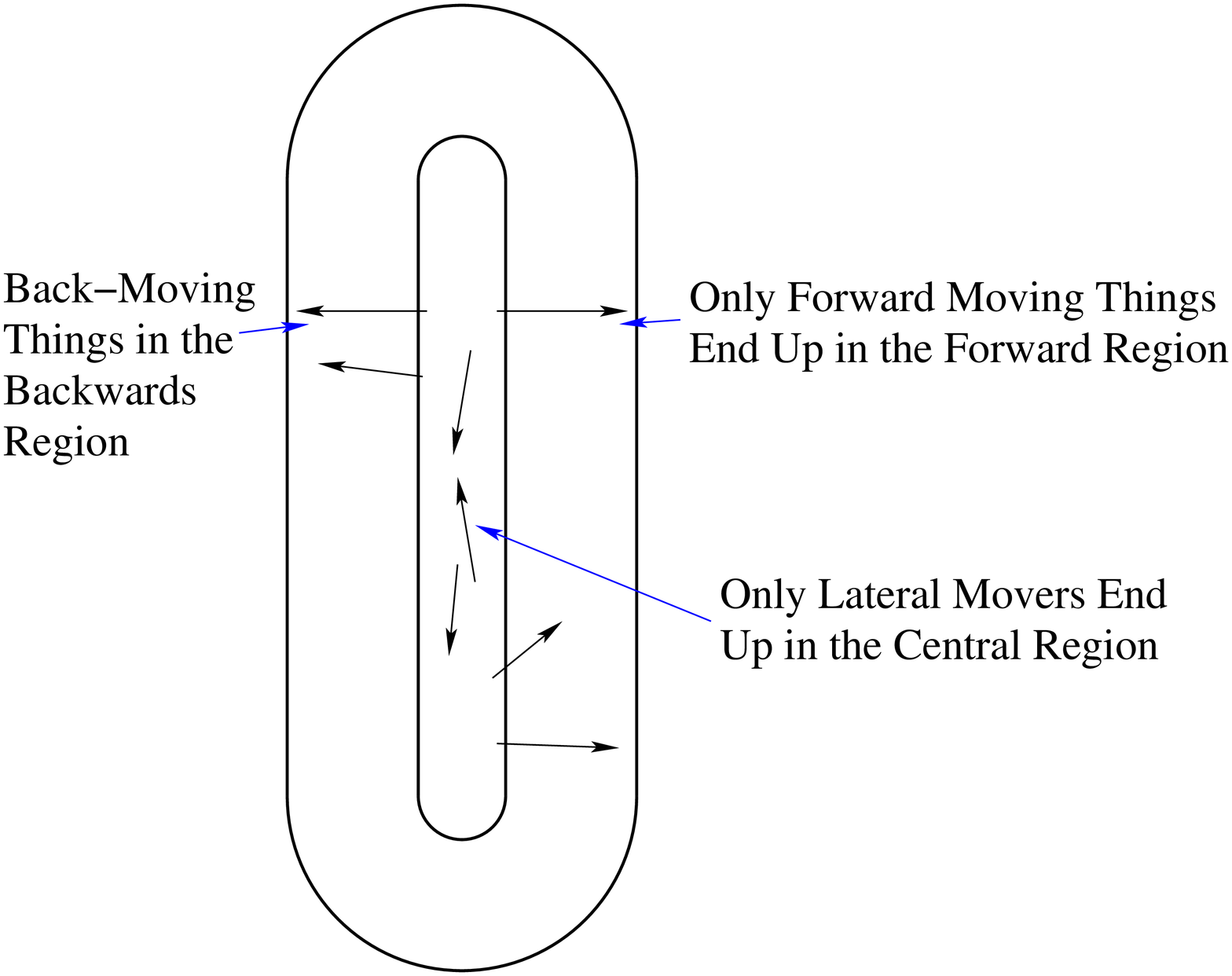}
\hfill \epsfxsize=0.4\textwidth\epsfbox{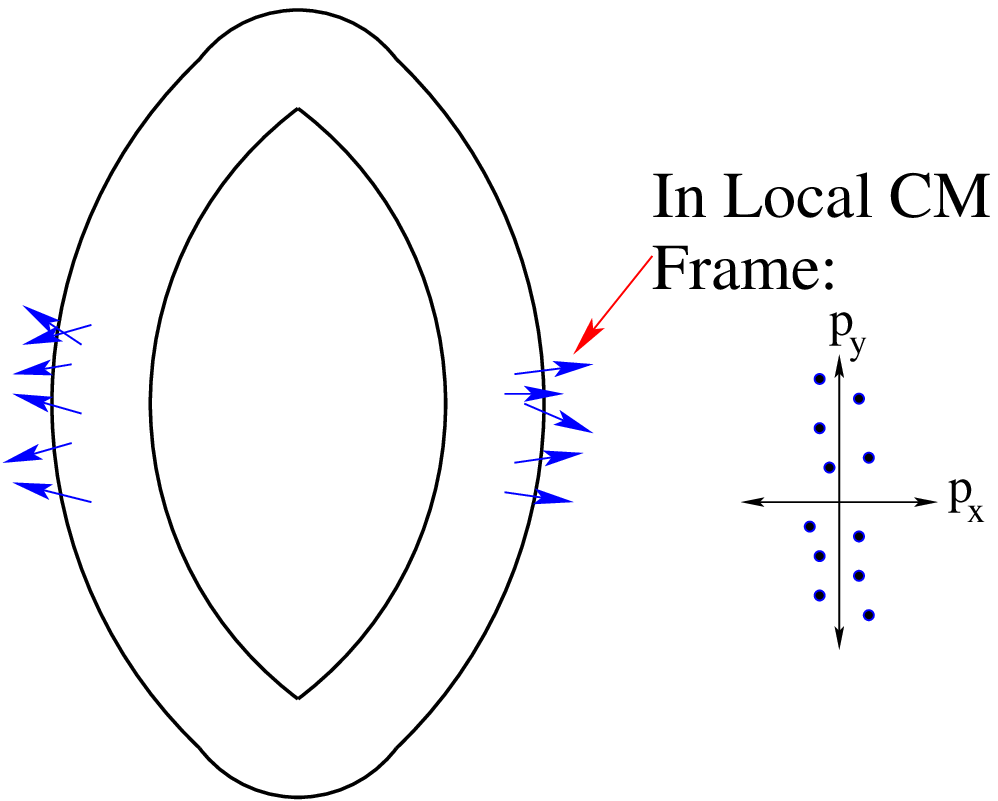}\hfill}
\caption{\label{fig2}Left:  momentum selection along the beam axis:
forward moving particles end up in front, backwards movers in back, and
only lateral movers remain in the middle.  Right:  the same applies in
the transverse plane.}
\end{figure}

This flat initial shape will quickly change, as the region containing
quark-gluon plasma expands into the space around it.  If the quarks and
gluons stream freely (as they would, at least initially, if the coupling
$\alphas$ were truly small), then ``momentum selection'' will
make the plasma locally highly anisotropic, see Figure \ref{fig2}.

In the absence of rescattering, the momentum distribution reaching
detectors would be isotropic, since the initial distribution of
particles was.  If rescattering is efficient, then the momentum
distribution relaxes to be isotropic, {\em locally} at each point, and
with respect to the local rest frame, which is a moving frame with
respect to the system as a whole.
For instance, in the righthand picture in Figure \ref{fig2},
the particles on the right side of the almond would redistribute to be
isotropic with respect to their (rightward moving) local rest frame.
Because the initial shape is anisotropic, there are more left and right
moving ``cells'' of plasma than up and down moving ``cells.''  The
$p_\perp^2$ summed over particles gets one contribution from the
$p_\perp^2$ of the particle relative to its cell, and one from the
$p_\perp^2$ of the cell.  The first is equal between $p_x^2$ and $p_y^2$
if rescattering isotropizes locally.  The second favors the direction
which was initially the skinny direction of the almond.  Therefore, the
momentum distribution of final particles will favor the initially thin
axis of the almond, {\em if} rescattering occurs.  This is called
elliptic flow, and it is observed to be nearly maximal in heavy ion
collisions \cite{elliptic_measure,elliptic_hydro}, that is, nearly at
the value obtained if rescattering is perfectly efficient.

It has been argued \cite{sQGP} that the elliptic flow observed in heavy
ion collisions is incompatible with the plasma being weakly coupled.
That is, perturbative treatments of the QGP \cite{molnar} seem
to be at a loss to
explain how it can show as much elliptic flow as it seems to.%
\footnote{%
    Inclusion of number changing processes may change this conclusion
    \cite{greinerxu}.  This is an
    interesting development, though we are concerned that those
    treatments do not include virtual (suppressing) corrections to the
    $2\leftrightarrow 2$ cross-section.}
However,
before discarding the idea of weak coupling, we should first make sure
that the weak coupling treatments were done correctly.  We
argue that weak coupling treatments to date leave out the
dominant physics, which has to be understood before such
strong claims can be made with confidence.

\section{Plasma instabilities}

Consider first what the traditional treatment of scattering in a plasma
is.  The particles in the plasma are considered to undergo
$2\leftrightarrow 2$ scattering.  The scattering rate has a (Coulombic)
soft divergence, $d\sigma \sim d^2 q_\perp / q_\perp^4$.  To find out
how much this isotropizes the plasma one must multiply by $q^2$ to get
the transport cross-section, but this is still log divergent:
\begin{equation}
\sigma_{\rm transport} \propto g^4 \int q^2 dq^2 \frac{1}{q^4} \, .
\end{equation}
This means that it is essential to include plasma corrections to the
scattering.  In equilibrium, this leads to a finite result
\cite{Gelis}:
\begin{equation}
\sigma_{\rm transport} \propto g^4 \int q^2 dq^2 \frac{1}{(q^2 + \Pi)^2}
\sim g^4 \int q_\perp^2 dq_\perp^2 \frac{1}{q_\perp^2 (q_\perp^2 +
m_{\rm D}^2)} \, .
\end{equation}
(Here $\Pi$ represents the self-energy and $m_{\rm D}^2$ is the Debye
mass squared.)
Plasma effects cut off the small angle scattering rate.  However, $\Pi$
was derived using the {\em isotropic} equilibrium result for the
self-energy.  For a nonequilibrium plasma,
one should recompute the self-energy for the nonequilibrium, {\em
anisotropic} particle distribution.  The relevant self-energies are
known \cite{ThomaStan}, but when they are inserted in the scattering
calculation, they lead to propagators which are singular at finite
momentum and give an apparently divergent answer for $\sigma_{\rm
transport}$ \cite{AMY_Boltzmann}.

We should examine the self-energy more carefully.  The leading ``hard
loop'' \cite{BraatenPisarski,Strickland} self-energy represents the
current induced by the coherent response of the plasma to an infrared
gauge field.  The simplest example to think about is a plasma
oscillation, which we will now describe for an abelian (E$\&$M) plasma.
Suppose that there is a spatially uniform $\E$ field in the plasma.
Each $+$ charge will be deflected into the direction of the $\E$
field, each $-$ charge will be deflected against it.  For a $+$ and $-$
charge at the same location and direction, the $+$ moves in the $\E$
direction and the $-$ against that direction, leading to a small
dipole.  Each individual deflection is small, but there are a lot of
particles and the deflections build up over time.  The individual
dipoles add up into an $\E$ field opposing the original one, which
soon cancels off the initial $\E$ field completely.  But the
particles keep following the deflected trajectories, so the dipoles
continue to grow, and the total $\E$ field reverses sign.  Then the
particles deflect back the other way, leading back to the starting
configuration.  The $\E$ field will oscillate back and forth in sign at
a characteristic frequency $\omega_{\rm pl}$ determined by the density
and deflectability of the charges in the plasma.

\begin{figure}
%
\centerline{
\epsfxsize=0.45\textwidth\epsfbox{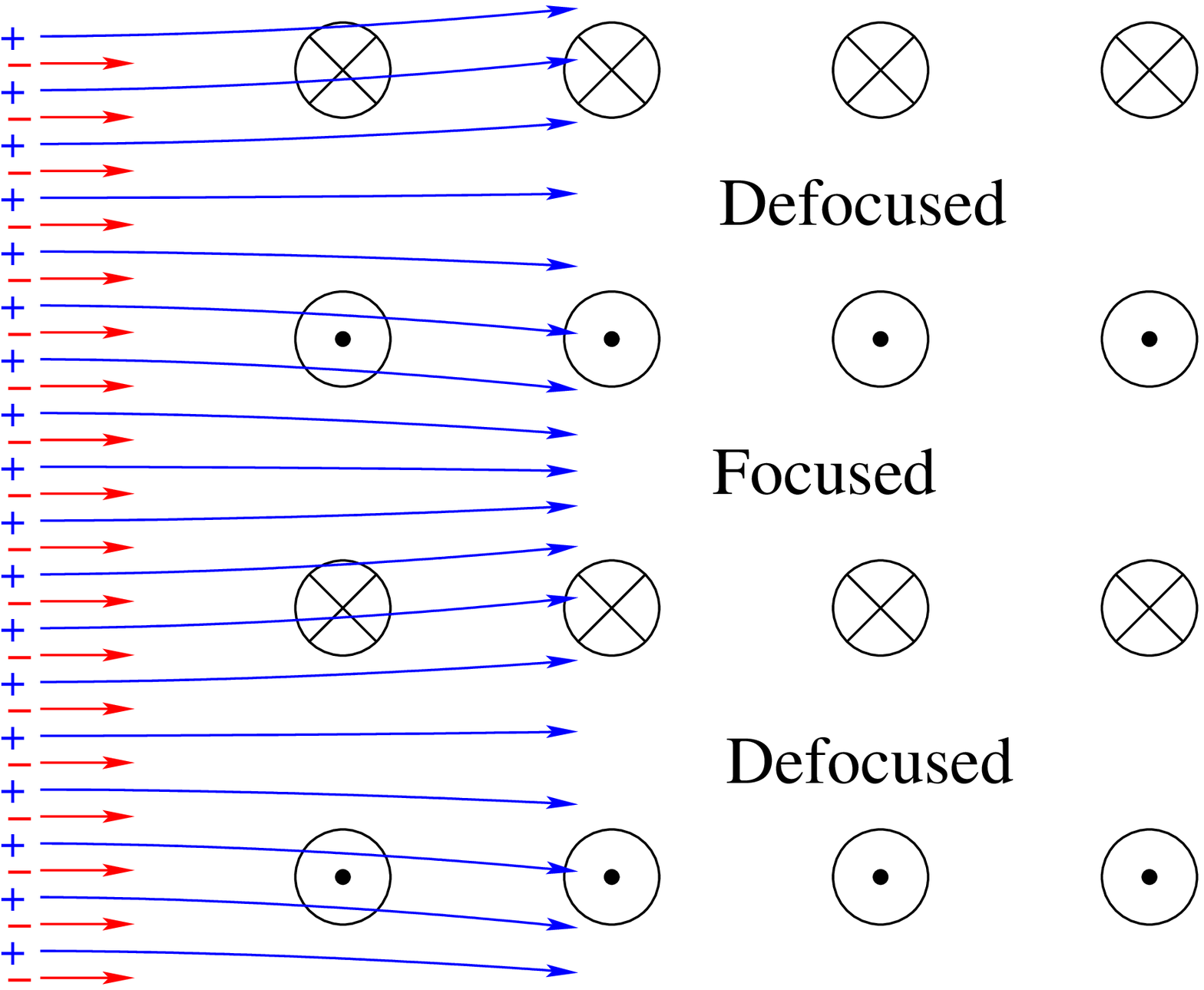} \hfill
\epsfxsize=0.45\textwidth\epsfbox{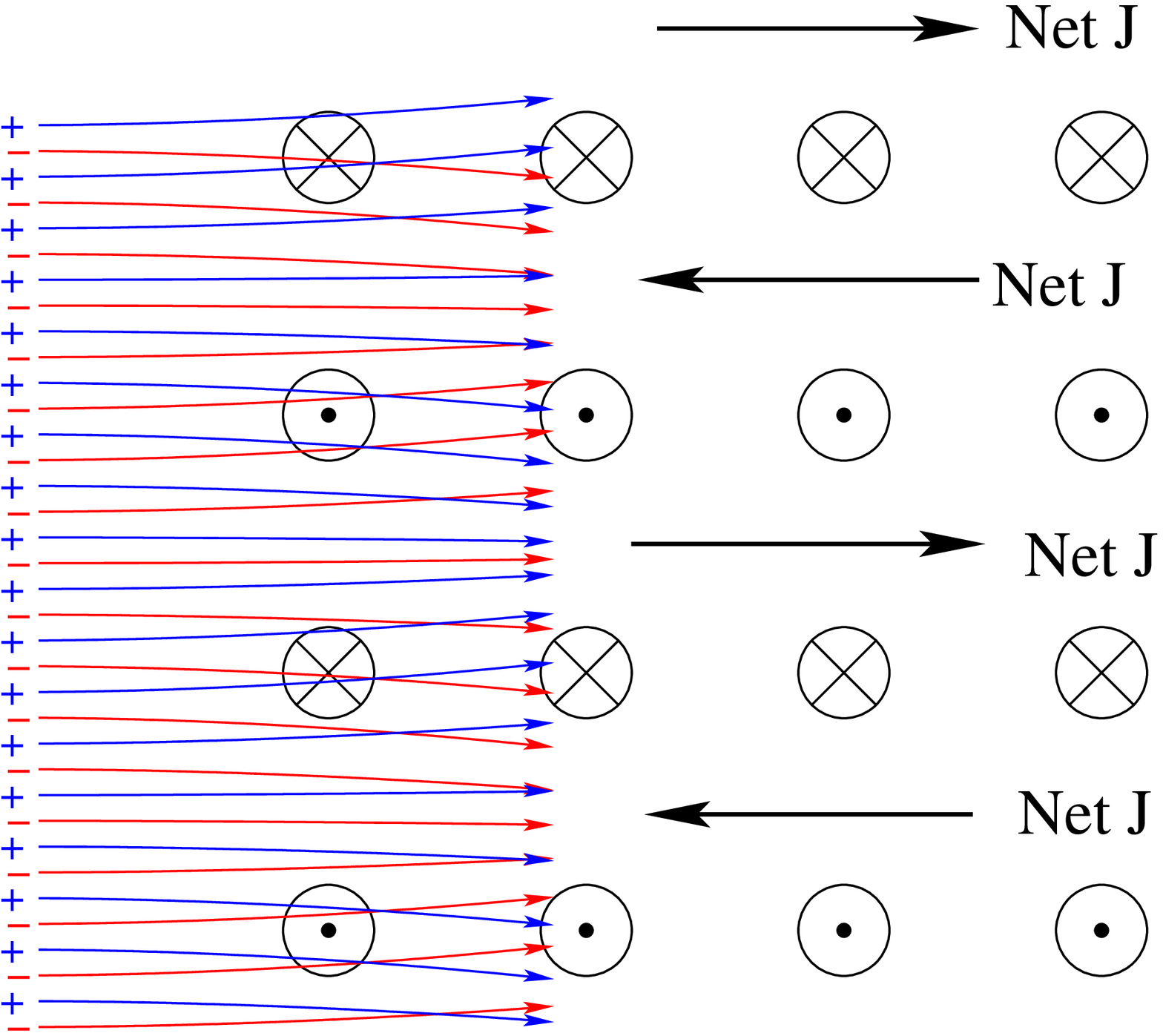}
}
\caption{\label{fig3} 
Left:  magnetic field and particle distribution which is unstable, and
the deflection of $+$ charges.  Right:
deflection of all charges, and induced currents.  The current is exactly
the one which supports the magnetic field.}
\end{figure}

Now consider a more complex and relevant example, shown in Figure
\ref{fig3}.  Suppose an anisotropic plasma has all charges flow along one
``beam'' direction.  Consider a seed magnetic field with $\k$ and
$\B$ orthogonal to the beam.  The particles deflect in the magnetic
field; the $+$ charges are focused where the $-$ charges are defocused,
leading to a net current.  The current is exactly the one which supports
the magnetic field which started the trouble.  Since the deflection
grows with time, the current will become large enough to generate and
even strengthen $\B$.  This will lead to greater charge deflection,
greater current, and greater $\B$.  This is an exponential
instability, called the {\sl Weibel instability}, known in the plasma
physics community since the 1950's \cite{Weibel}.  Though we discussed
the case of maximal instability, the same process is present whenever
the plasma is anisotropic.  The time scale for this
instability to grow is short:  if typical momenta are $O(T)$ and the particle
density is $O(T^3)$, but the momentum distribution is anisotropic, the
growth rate of the instability is $\gamma \sim gT$, to be compared with
the large angle particle scattering rate (in equilibrium) of $\Gamma
\sim g^4 T$.  Therefore the $\B$ fields become large on a scale
shorter than any scattering time scale in the plasma.  In fact, it is
generally faster than {\it any} dynamical time scale in the plasma, or
the system age \cite{ALM}, and {\em must} be considered in understanding
the evolution of the plasma.

\section{Kinetic theory treatment}

We would like to address plasma instabilities in the context of the full
QGP evolution at realistic coupling.  That is a big task.  We take
instead the warmup task of understanding behavior in the limit of weak
coupling $\alphas \ll 1$, some time after the collision when the
plasma has become anisotropic, and working only in a local patch of
plasma which is statistically spatially uniform.  The key is that the
particle density $n$ becomes smaller than the nonperturbative size,
$n \ll p^3/\alphas$, with $p$ the typical momentum of a hard particle.
Therefore the screening scale $m^2 \sim \alpha n/p$ becomes $m\ll p$;
there is a separation of scale between the wave number of the unstable
modes and of the dominant excitations in the plasma.  This allows a
separation of the degrees of freedom.  The soft modes will achieve large
occupancy due to the instability, and may therefore be treated as
classical fields.  The dominant modes have large momentum and may
therefore be treated as particles.  This allows a Vlasov equation
treatment.

In equilibrium, an electric field polarizes the plasma, but a magnetic
field merely rotates the isotropic distribution of momenta, which does
not induce a net current.  For an anisotropic plasma, though, the
distribution of particles, 
\begin{equation}
\Omega(\v) \equiv \frac{1}{\int d^3 p f(p)/p}
	\int \frac{d^3 \p}{p} f(\p) \delta(\hat\p-\v)
\end{equation}
is not isotropic.  ($\Omega(\v)$ represents the number of particles
moving in the $\v$ direction, weighted by their ``bendability'' $1/p$
and normalized to average to 1 over angles.)  This means that any
electromagnetic field can induce a current.  In particular, defining the
net color of particles moving in the $\v$ direction at point $\x$ 
as $W^a(\x,\v)$, the joint equations for $W$ and $A_\mu$ are,
\begin{eqnarray}
D_t W^a(x,\v)&=& - \v \cdot \D W^a(x,\v)
	+m_{\infty}^2\;\mbox{Source} \, , \\
\mbox{Source} & = &  2\Omega({\bf v}) \v\cdot \E^a
	-\E^a \cdot \frac{\partial}{\partial \v} \Omega({\bf v})
	-F^a_{ij} v_i \frac{\partial \Omega({\bf v})}{\partial v_j} \, ,
\\
D_\mu F^{\nu \mu} & = & J^\nu = \int_{\bf v} v^\nu\; W({\bf v}) \, .
\end{eqnarray}
The first equation describes the (covariant) free propagation of
particles, $v^\mu D_\mu W=0$, modified by the induced current from
polarizing the mean (colorless) distribution of particles.  The
parameter $m_\infty^2 = g^2 C_f \int d^3 p f(p)/p$ equals $m_{\rm
D}^2/2$ in equilibrium and gives the polarizability of the medium.  The
second line shows how much $\E$ and ${\bf B}$ fields polarize the
medium.  The last line is the Yang-Mills equation with $J$ determined by
$W$.

These equations can be solved two ways.  They can be handled
analytically in the context of weak field perturbation theory.  This has
been pursued quite far \cite{Mike2,ALM}.  The conclusion is that any
anisotropy causes exponential growth, and $O(1)$ anisotropy ($\Omega$
deviating by $O(1)$ amounts from being isotropic) leads to an
exponentiation time $\gamma \sim m_\infty$.  At any time later than the
formation time of the plasma, multiple exponentiation times have
occurred, and one must deal with the fully nonlinear equations.  These are
not tractable analytically, and we must use the other technique for
solving these equations:  numerical implementation.

The classical Yang-Mills equations can be solved nonperturbatively in
real (Minkowski) time on the lattice via existing techniques
\cite{Ambjorn}.  The $W$ fields have been added to such a lattice
simulation for an isotropic plasma in \cite{BMR}.  New difficulties
emerge in this treatment, because $W^a$ is a function of $\v$ as well as
$\x$; both must be discretized somehow to make the system finite.  Ref.\
\cite{BMR} does so by expanding functions of $\v$ in spherical harmonics
and truncating at a finite $\ell_{\rm max}$.  It is also possible to
discretize the sphere directly \cite{Mike3}.  Here we will present
results using the $\ell_{\rm max}$ cutoff technique, extended to
anisotropic $\Omega$ \cite{linear1}.  We work in SU(2)
rather than SU(3) for simplicity; we expect the qualitative behavior to
be the same, and the simulations remain at the level of understanding
the gross features rather than the quantitative details at this time.

\section{Numerical results}

We summarize results obtained jointly with Peter Arnold and Larry Yaffe,
which have recently been presented in greater detail
\cite{linear1,linear2}.

\begin{figure}
\centerline{\epsfxsize=0.45\textwidth\epsfbox{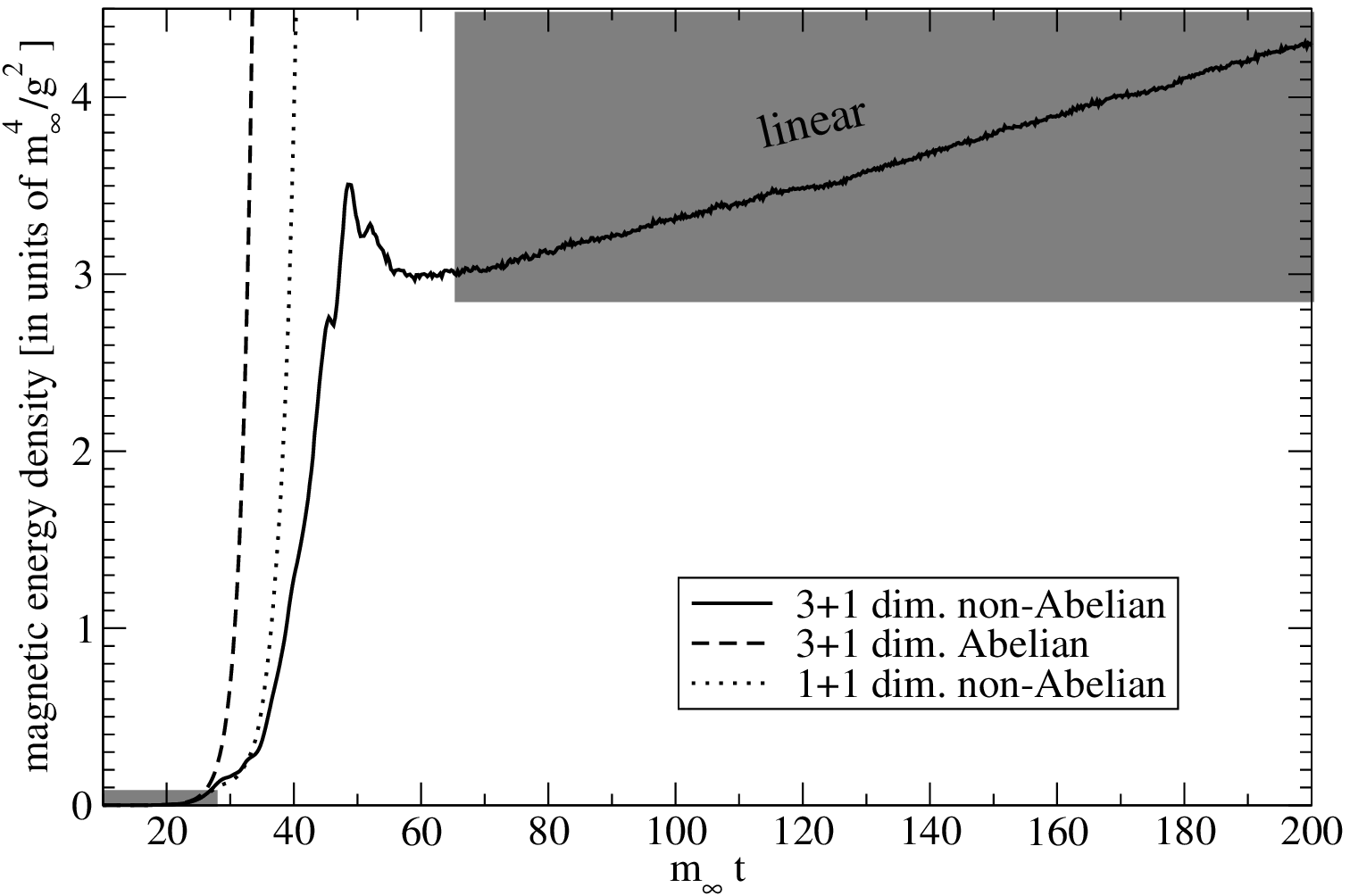} \hfill
\epsfxsize=0.45\textwidth\epsfbox{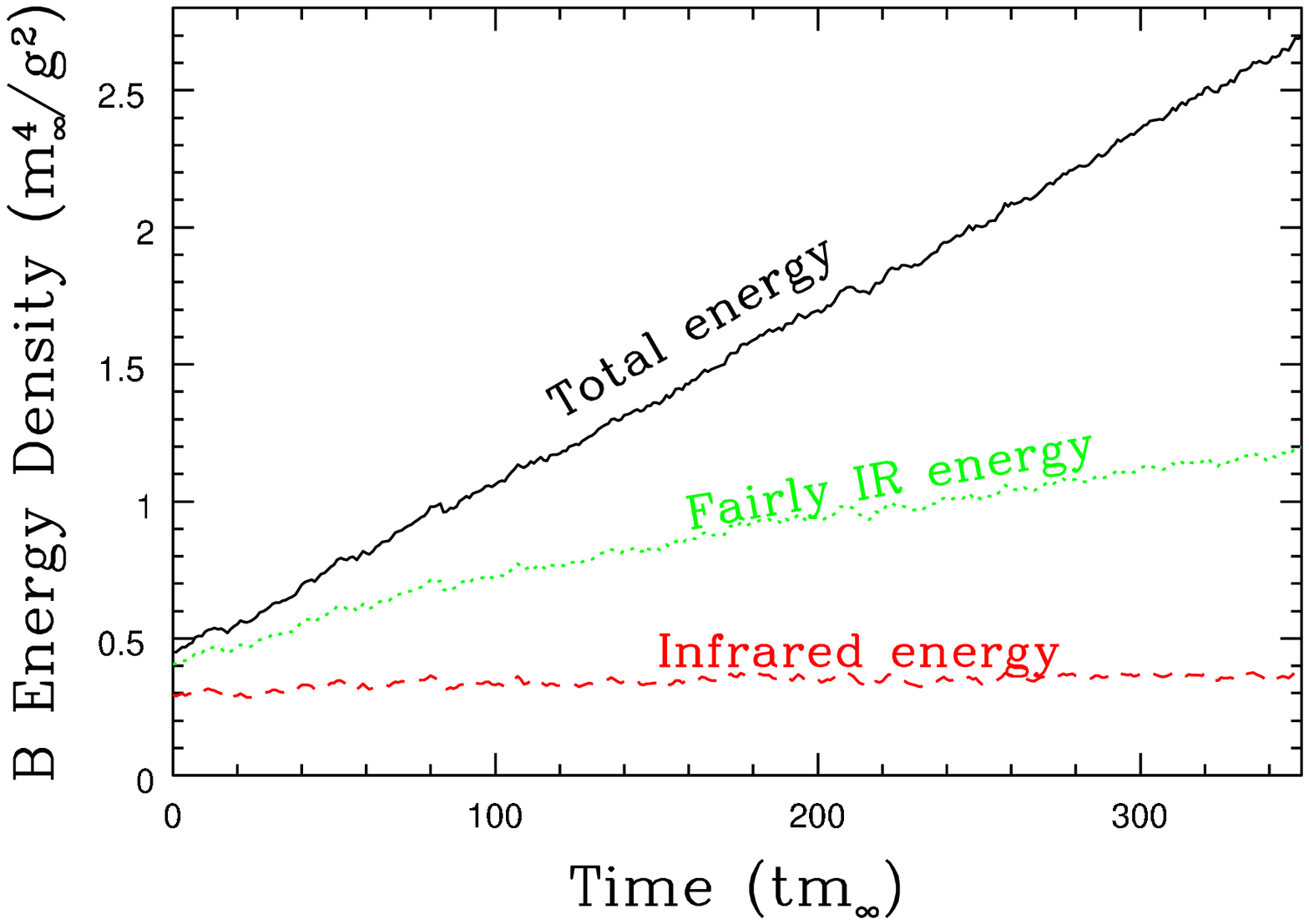}}
\caption{\label{fig4} Left:  energy growth from small seed fields is
exponential at first but becomes linear when the fields become
nonperturbative.  Right:  for nonperturbative seeds, the growth is always
linear; but the energy gain is in the higher $\k$ modes only, as
revealed by smearing.}
\end{figure}

If the gauge fields are initially small, they grow exponentially, as
expected perturbatively, until they are nonperturbatively large.  At
this point, the growth in the gauge field energy changes over to linear
behavior, see Figure \ref{fig4}.  What does this behavior represent?  By
``smearing'' the fields to remove high $\k$ components, we can
demonstrate (same figure) that the infrared fields do {\em not} grow any
more, but the energy goes into ever more ultraviolet degrees of
freedom.

This behavior is confirmed by looking at the power spectrum of the
electric and magnetic energy in Coulomb gauge.  In a system of
quasiparticles, the occupancy would obey,
\begin{equation}
f(k) = \frac{\langle E^2(k) \rangle}{2(N_{\rm c}^2-1)k} 
	= \frac{\langle k^2 A^2(k)\rangle}{2(N_{\rm c}^2-1)k} \, ,
\end{equation}
where $2(N_{\rm c}^2-1)$ just counts degrees of freedom in $E$ and $B$.
One can therefore define $f_{E}$ and $f_{B}$ as the occupancy as
determined by the electric and magnetic fields by imposing this
expression.  This is just a way of parameterizing the power spectrum of
$E$ and $A$; but if the degrees of freedom are actually behaving as
light quasiparticles, we should observe $f_B \simeq f_E$.

\begin{figure}
\centerline{\epsfxsize=0.45\textwidth\epsfbox{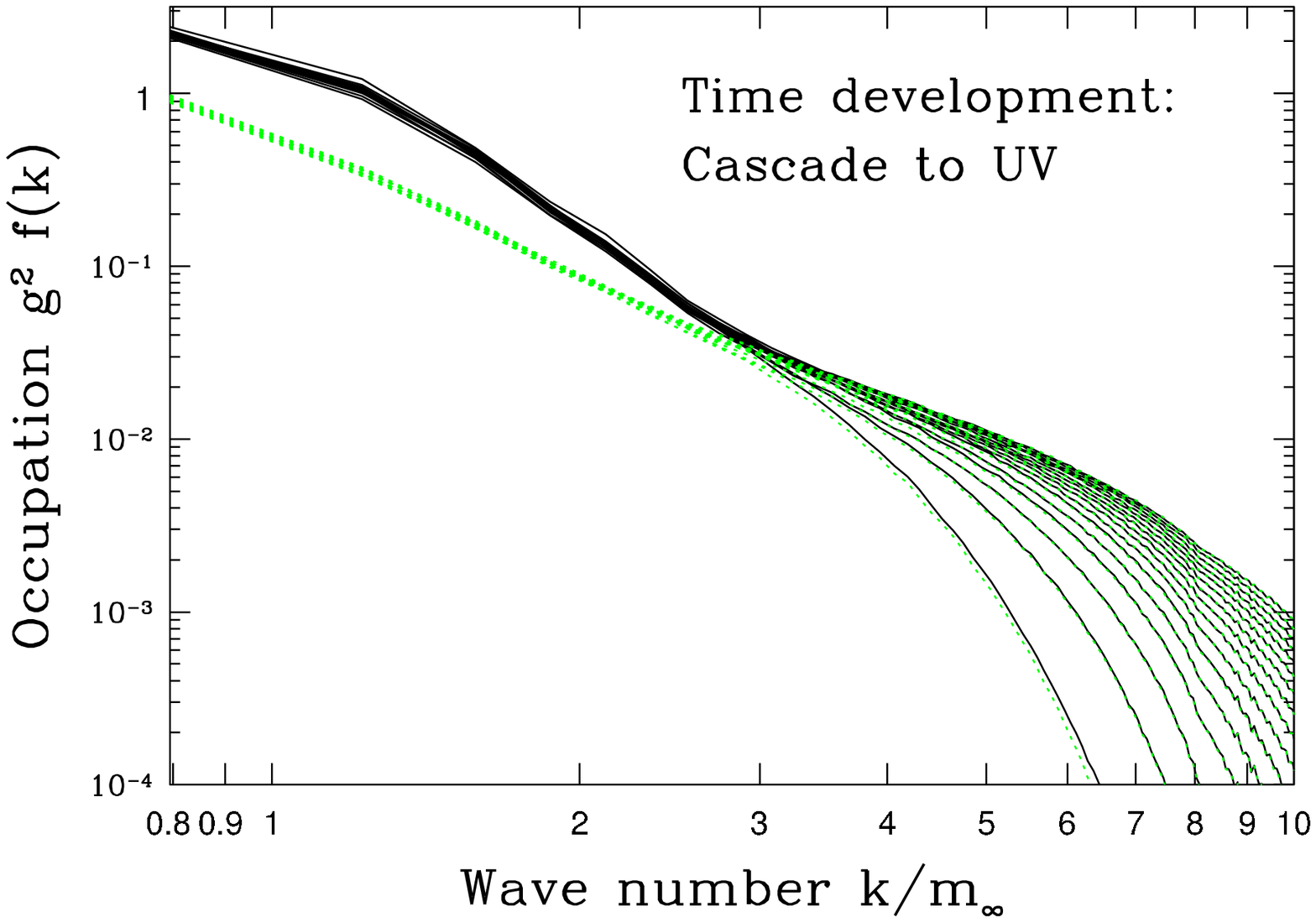} \hfill
\epsfxsize=0.45\textwidth\epsfbox{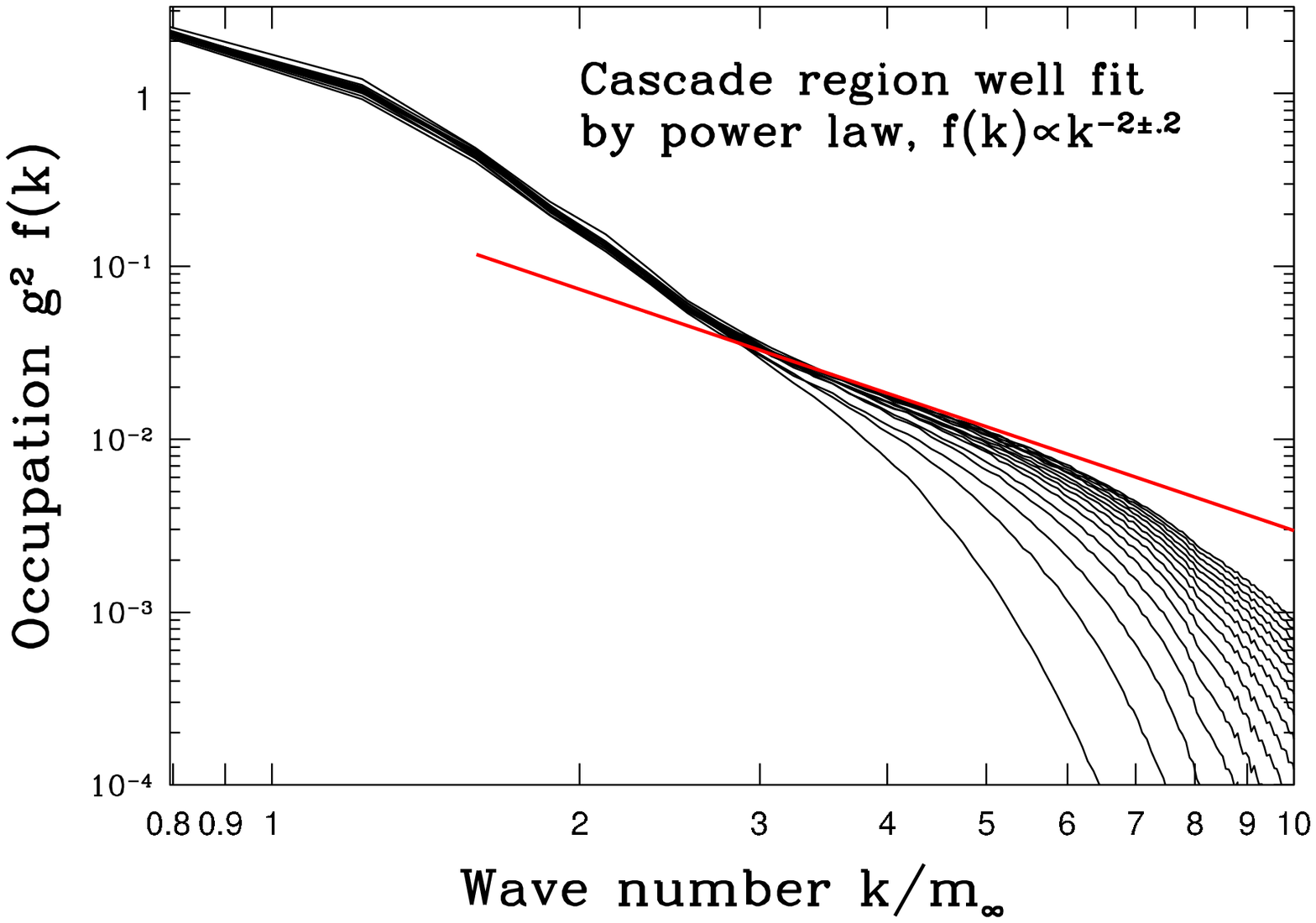}}
\caption{\label{fig5} Left:  Coulomb gauge spectrum from $E$ (lower at
left) and $A$ (higher at left) fields, at a series of times
from early (bottom at right) to late (top at right).  Right:  the same,
with a suggestive power law fit superimposed.  It appears that $f(k)
\propto k^{-2}$.}
\end{figure}

Results for $f_E$ and $f_B$ are shown in Figure \ref{fig5}.  The IR
contains nonperturbatively large fields which are quasistationary in
time.  In the UV, the behavior is that of quasiparticles, which become
more numerous with time, growing towards a steady state
power-law spectrum with a spectral index of approximately $f \propto
k^{-2}$.

While the IR behavior is quasistationary, it is dynamical, not
quasistatic.  Chern-Simons number diffuses (not shown), indicating that
nonperturbative physics is involved and that there is no long time scale
coherence to the soft gauge field configuration.

\section{Conclusions}

The expansion of the QGP in a heavy ion collision should lead to a
locally anisotropic system.  Within the weak coupling expansion, this
implies that plasma instabilities should develop.  Plasma instabilities
imply a transfer of energy from the ``hard'' typical excitations to
large long wavelength ``soft'' nonabelian magnetic fields.  The
subsequent evolution of these soft fields requires nonperturbative tools
to uncover.  In an $\alphas\ll 1$ treatment where the separation between
hard and soft scales is parametrically large, we find an intriguing
cascade phenomenon.  Energy is taken from the hard fields into the soft
fields.  However, nonabelian interactions between soft fields keep the
would-be unstable fields in a quasisteady state.  Instead the energy
cascades into classical fields with wave numbers $k$ larger than the
unstable field scale, $m_\infty$.  The cascade towards the ultraviolet
develops with a power law spectrum $f \propto k^{-\alpha}$ with $\alpha
\simeq 2$.  This value of the spectral index implies that the cascade
particles do not dominate the screening and do not serve as the dominant
source of scattering events.

We do not fully understand the implications of these results for
thermalization of the QGP even in the case of weak coupling, much less
at realistic couplings for current experiments.  However, the physics is
rich and intriguing and deserves further study.

\end{document}